\documentstyle[a4wide,epsf]{article}

\begin{document}

\newcommand{\lsim}{\mbox{\raisebox{-.9ex}{~$\stackrel{\mbox{$<$}}{\sim}$~}}}
\newcommand{\gsim}{\mbox{\raisebox{-.9ex}{~$\stackrel{\mbox{$>$}}{\sim}$~}}}

\begin{center}
{\large\bf  Locked Quintessence and Cold Dark Matter}

\bigskip

{\large  Minos Axenides$^*$ }
\bigskip

$^*${\it Institute of Nuclear Physics, National Center for
Scientific Research
`Demokritos',\\
Agia Paraskevi Attikis, Athens 153 10, Greece}

\begin{abstract}
A supersymmetric hybrid potential model with low energy
supersymmetry breaking scale ($ M_{S}\sim 1-10 Tev$) is presented
for both dark matter and dark energy. Cold dark matter is
associated with a light modulus field ($\sim 10-100 Mev$)
undergoing coherent oscillations around a saddle point false
vacuum with the presently observed energy density ($\rho_{0} \sim
10^{-12} eV^{4}$). The latter is generated by its coupling to a
light dark energy scalar field ($ \sim 10^{-18} eV $) which is
trapped at the origin ("locked quintessence"). Through naturally
attained initial conditions the model is consistent with cosmic
coincidence reproducing LCDM cosmology. An exit from the cosmic
acceleration phase is estimated to occur within some eight Hubble
times.
\end{abstract}

\end{center}

\section{Introduction}

There is a growing observational evidence to the fact that we live
in a spatially flat Universe ($\Omega_{tot}\approx 1$) in a state
of cosmic acceleration \cite{wmap, sdss, 2dF, SN}. Most of its
content, by weight ($\Omega_{tot}-\Omega_{bar}\sim 0.96 $), cannot
be accounted for by the standard model of particle physics. It is
believed to be associated with an invisible sector of Matter and
Energy of, remarkably, almost equal energy density in a cosmic
coincidence. Dark Matter( $\Omega_{DM}\sim 0.3$), responsible for
the growth of structure in our Universe, is believed to be
non-baryonic in nature with small free streaming length behaving
as a non-relativistic gas (Cold Dark Matter-CDM). It is typically
associated with weakly interacting massive particles (WIMPs) such
as axions, axinos, neutralinos, gravitinos, string moduli and
others\cite{sahni}.

Dark Energy ($\Omega_{DE}\sim 0.7$), on the other hand, is
probably a homogeneous perfect fluid component($p\sim w \rho$)
with negative pressure ($w < \ - \frac{1}{3}$)  giving rise to the
observed cosmic acceleration(for a review see \cite{review}). In
its most popular version it is attributed to the Cosmological
Constant ($w=-1$) whose value must be fine tuned to an unprecedent
degree to be in accordance with the observational data
($\frac{\Lambda}{8\pi G}\sim 10^{-47}$). The emerging LCDM
Cosmology, although  economical and succesful  is not lacking of
theoretical shortcomings. Indeed a constant vacuum energy
inevitably leads to eternal accelerated expansion , technically
implying the presence of causal horizons and hence non-existence
of well defined in and out states in the formulation of the
underlying quantum theory such as String theory\cite{horizons}.

Alternative scenarios employ dynamical scalars , such as
Quintessence fields ($ -1 < w <-1/3 $) \cite{quint} which possess
time varying energy density as they roll down their monotonically
decreasing potential energies . They typically predict an exit
from the present accelerating phase. Eventhough these models
dispense with the theoretical problems of the Cosmological
Constant scenario they dont lack unnatural fine tunnings\cite{KL}
associated typically with both their initial conditions, present
value and/or their small mass ($M_{Q}\sim 10^{-33} $ eV). In the
context of supergravity theories such a light field is difficult
to be understood because the flatness of its potential is lifted
by excessive supergravity corrections or due to the action of
non-renormalizable terms, which become important at displacements
of order $M_P$.

Cosmic acceleration in the very early universe has been
extensively studied in supersymmetric hybrid models\cite{hybrid}.
There the vacuum energy density required in order to generate the
necessary number of e-foldings is fed into the slow rolling
inflaton through its coupling to a second scalar field the
"waterfall" which is kept trapped along the inflaton track. In a
fast-roll variation of this scenario, also dubbed "locked
inflation"\cite{dvali, ours1}, the inflaton field undergoes rapid
coherent oscillations  around  its "Saddle point" vacuum before it
is displaced away from it, prolonging consequently the
inflationary phase.

Interestingly it has been known for quite a while that coherent
oscillations of massive (pseudo)scalar weakly interacting
particles, such as the axion can mimic Cold Dark Matter ($w=0$)
\cite{turner}. We have recently produced an interacting model that
realizes LCDM Cosmology by putting these two ingredients together
in the very late universe \cite{ours2}. Other interacting models
for Dark Matter and Dark Energy can be found in Ref.~\cite{other}.

Our  model is a given by a standard Supersymmetric Hybrid
Potential with only two characteristic energy scales : the Planck
Mass ($M_{Pl}\approx 10^{19} GeV$) and a low energy SUSY breaking
scale ($M_{S}\approx M_{3/2}\approx 1 TeV$). We assume that the
dark matter particle is a modulus $\Phi$, corresponding to a flat
direction of supersymmetry. The modulus field is undergoing
coherent oscillations, which are equivalent to a collection of
massive $\Phi$--particles ($ M_{\Phi}=\frac{M_{S}^{2}}{M_{Pl}}\sim
10-100 MeV $), that are the required WIMPs. A second scalar field
($ \Psi $)interacts with ($\Phi$) in a standard way
($\lambda\Phi^{2}\Psi^{2}$). This can be thought of as our
quintessence field and it corresponds to a flat direction lifted
by non-renormalizable terms. Even though the $\Psi$--field is a
light scalar ($M_{\Psi}\approx \frac{M_{S}^{3}}{M_{Pl}^{2}}
\approx 10^{-18}$ eV) , it is much more massive than the $m_Q$
mentioned above, so as not to be in danger from supergravity
corrections to its potential\cite{book,sugra}. Our quintessence
field is coupled to our dark matter in a hybrid manner, which is
quite natural in the context of a supersymmetric theory. Due to
this coupling, the oscillating $\Phi$, keeps $\Psi$ `locked' on
top of a potential hill, giving rise to the desired dark energy.
When the amplitude of the $\Phi$--oscillations decreases enough,
the dark energy dominates the Universe, causing the observed
accelerated expansion as dictated by the cosmic coincidence.
Within some eight Hubble times , when the oscillating amplitude
falls below the width of the saddle, the `locked' quintessence
field is released and rolls down to its global minimum. The system
reaches the true vacuum and accelerated expansion ceases.

We assume a spatially flat Universe, according to the WMAP
observations \cite{wmap}. We use natural units such that
\mbox{$\hbar=c=1$} and Newton's gravitational constant is
\mbox{$8\pi G=M_{Pl}^{-2}$}, where \mbox{$M_{Pl}= 10^{18}$GeV} is
the reduced Planck mass.

\section{ A Supersymmetric Hybrid Model}\label{model}

Consider two real scalar fields $\Phi$ and $\Psi$ interacting
through  a hybrid type of potential of the form\cite{ours2}
\begin{equation}
V(\Phi,\Psi)=\frac{1}{2}m_\Phi^2\Phi^2+
\frac{1}{2}\lambda\Phi^2\Psi^2+\frac{1}{4}\alpha(\Psi^2-M_{s}^2)^2,
\label{V}
\end{equation}
where ($\lambda\leq 1$) and
($\alpha=\frac{M_{s}^{4}}{M_{Pl}^{4}}$). Dark Matter is associated
with $\Phi$ and Dark Energy with $\Psi$. All parameters are
expressed in terms of two fundamental energy scales: the Planck
mass ($M_{Pl} \sim 10^{18}$ GeV) and the Susy breaking scale which
is also taken to be the Gravitino mass ($M_{s} \sim m_{3/2}\sim 1
TeV$). They should be considered in the framework of gauge
mediated supersymmetry\cite{giudice}. Standard features of the
potential which is  depicted in the
  figure are:
\begin{itemize}
    \item
    Two global minima $(\Phi,\Psi)=(0,± M_{s})$ with an
    unstable saddle point at $(\Phi,\Psi)=(0,0)$
    \item
    $\Psi$ possesses a $\Phi$ dependent curvature
    $(m_\Psi^{\rm eff})^2=\lambda\Phi^2-\alpha M_{s}^2)$
    with a width of
    \begin{equation}
    \Phi_{w}=
    \frac{1}{\sqrt{\lambda}}\frac{M_{s}^{3}}{M_{Pl}^{2}}
    \label{width}
    \end{equation}
    \item
    Cosmic Coincidence ($\Omega_{DE} -\Omega_{DM}\sim
    {\cal O} (1)$)at present Hubble time demands small scalar masses :
    $m_{\Phi}\sim\frac{M_{s}^{2}}{M_{Pl}}\approx 10-100 MeV $  and
    $m_{\psi}\sim \frac{M_{s}^{3}}{M_{Pl}^{2}}\approx 10^{-18}$eV.
    The latter is conceivable to be due to accidental
    cancellations in the K\"{a}hler potential or some other
    accidental symmetry protecting $m_{\Psi}$.
    \end{itemize}
\begin{center}
\begin{figure}
\label{fig}
\begin{center}
\leavevmode
\hbox{%
\epsfxsize=3in \epsffile{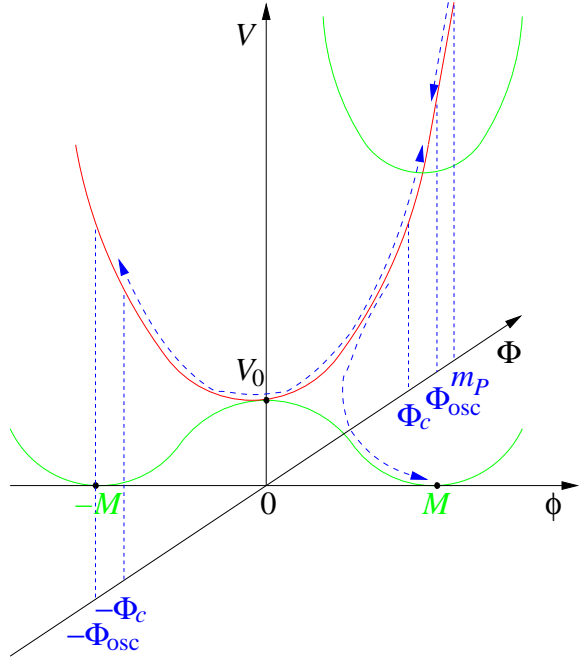}}
\end{center}
\caption{\footnotesize Illustration of the scalar potential
$V(\Phi,\Psi)$. Originally, \mbox{$\Phi\sim M_{Pl}$} and
\mbox{$\Psi\simeq 0$}. The field $\Phi$ begins oscillating with
amplitude $\Phi_{\rm osc}$. Due to the expanding Universe its
energy gets diluted until it reaches \mbox{$\bar{\Phi}_{\rm
end}\sim\Phi_w$}, when the system departs from the saddle and
rolls toward the minimum at \mbox{$(\Phi=0, \Psi=\pm M_{s})$}. }
\end{figure}
\end{center}
  Our physical system acts as a two component perfect fluid
  with energy density $(\rho_{tot}= \rho_{\Phi} + \rho_{\Psi})$ which gets diluted
  as the universe expands.
  When the system finds itself rolling at $\Phi\geq \Phi{w}$
  it is energetically favorable for $\Psi$ to be
  trapped in its origin $\Psi \cong 0$. $\Phi $ performs coherent oscillations in a quadratic
  potential
\begin{equation}
V(\Phi,\Psi=0)=\frac{1}{2}m_\Phi^2 \Phi^2 + V_{0}. \label{over}
\end{equation}
around a saddle point false vaccuum with energy density given by
\begin{equation}
V_{0} =\frac{1}{4}   \alpha M_{s}^4 \sim \frac{M_s^8}{M_{Pl}^{4}}
\sim 10^{-120} M_{Pl}^{4} \label{V0}
\end{equation}
 the observed present vacuum energy density.
 It is associated with
 the DE condensate $\Psi$ acting as an effective cosmological constant, i.e.
  behaving  as a perfect fluid component
 with an equation of state $(p=- V_{0})$.
 In the high temperature
 phase  the energy density is dominated by the kinetic energy of
 the $\Phi$ oscillations which behaves as a pressureles
 non-relativistic  component of a collection of massive particles,
 hence Cold Dark Matter(CDM), which is
 given by :
\begin{equation}
   \rho_\Phi = \frac{1}{2}\dot{\Phi}^2+\frac{1}{2}m_\Phi^2\Phi^2,
\label{rPhi}
\end{equation}
where the dot denotes derivative with respect to the cosmic time
$t$. The model therefore identifies the following cosmic phases:
\begin{description}
    \item[\underline{CDM domination}]
     The overall density is dominated by the coherent oscillations of $\Phi$
     in Eq.~(\ref{rPhi}), when the oscillation amplitude is larger than
\begin{equation}
\Phi_\Lambda \sim\frac{\sqrt{\alpha}\,M^2}{m_\Phi}
\sim\left(\frac{m_\Psi}{m_\Phi}\right)M_{s}\sim
\frac{M_{s}^{2}}{M_{Pl}} \label{PhiL}
\end{equation}
They behave as a collection of non-relativistic particles whose
energy gets diluted accordingly as  ($\rho_{\Phi}\propto R^{-3}$)
    \item[\underline{Locked Quintessence}]
    The energy density is dominated by the Saddle Point Vacuum of
    eq.(\ref{V0}) for the range of $\Phi$ amplitude oscillations
    \begin{equation}
    \Phi_{w} < \Phi_{0} < \Phi_{\Lambda}
    \end{equation}
 The characteristic time scale that $\Phi$ spends on the saddle
($ \Phi_{0} < \Phi_{w}$) is ($\Delta t_{w} \sim
\frac{\Phi_{w}}{m_{\Phi} \Phi_{0} }$). As long as it is smaller
than the time scale ($\Delta t_{\Psi} \approx \frac{1}{m_{\Psi}}$)
it takes for $\Psi$ to start to roll away from the top of the hill
rapid coherent oscillations of $\Phi$ persist (Locked
Quintessence). The effect is present due to the ratio of masses
chosen ($\frac{m_{\Phi}}{m_{\Psi}}\approx
\frac{M_{Pl}}{M_{s}}\approx 10^{16}$). A (quasi) de Sitter
expansion phase sets in with \mbox{$a\simeq a_0\exp(H_0\Delta
t)$}, where \mbox{$\Delta t=t-t_0$} and
\mbox{$H_0\simeq\sqrt{V_0}/\sqrt{3}\,M_{Pl}=$ constant}. For the
oscillating $\Phi$ we have
\mbox{$\Phi\propto\sqrt{\rho_\Phi}\propto a^{-3/2}$}. We can thus
obtain an estimate of the length of the cosmic acceleration phase.
    \item[\underline{Post-Acceleration Phase}]
Our two fluid system will release its stored vacuum energy when
$\Psi$ will start to roll away from the top of the hill away from
its present false vacuum state into its future true vacuum of zero
energy density when ($\Phi_{0}=\Phi_{w}=\Phi_{\Lambda}$)
\begin{eqnarray}
 & \Phi_w\simeq\Phi_\Lambda\exp(-\frac{3}{2}H_0\Delta t_w) &
\Rightarrow\qquad \Delta t_w\simeq\frac{2}{3}
\left[\,\ln\left(\frac{M_{Pl}}{M_S}\right)
+\ln\sqrt{\lambda}\,\right]H_0^{-1},
\end{eqnarray}
We see that the period of acceleration may last up to wight Hubble
times (e-foldings) depending on the value of $\lambda$.
\end{description}
\section{ Dark Matter and Dark Energy Requiremenets}
\begin{itemize}
    \item
     Coherent Oscillations of the modulus $\Phi$ field in a
     quadratic potential behave as a collection of massive
     non-relativistic particles. In order
     that we may identify them with a realistic CDM component
     ($\Omega \approx \frac{1}{3}$) they must persist until today
     with the $\Phi$ quanta not having decayed , namely
     satisfying
     \begin{equation}
     \Gamma_\Phi < H_{0}\;, \label{GPhi}
     \end{equation}
     where \mbox{$H_0\sim\sqrt{\rho_0}/M_{Pl}$} is the Hubble parameter at
     present. Using that \mbox{$\Gamma_\Phi\sim g_\Phi^2m_\Phi$} we
     find the bound
     \begin{equation}
     m_\Phi\leq 10^{-20}M_{Pl}\;, \label{mPhibound}
     \end{equation}
     where we used that the coupling $g_\Phi$ of $\Phi$ with its decay
     products lies in the range \mbox{$\frac{m_\Phi}{M_{Pl}}\leq
     g_\Phi \leq 1$}, with the lower bound corresponding to the
     gravitational decay of $\Phi$, for which \mbox{$\Gamma_\Phi\sim
     [m_{\Phi}^3]/M_{Pl}^2$}. We may conclude  that $\Phi$ has to be a
     rather light field with mass \mbox{$\lsim$ 10-100 MeV}.
     \item
       We must require that our dark matter field $\Phi$
       should not decay into $\Psi$-particles, through their mutual coupling,
       until the present time either perturbatively  (\mbox{$\Phi\rightarrow\phi\;\phi$})
       or non-perturbatively through parametric resonance.
       The perturbative condition reads
       \begin{equation}
       \Gamma_{\Phi\rightarrow\phi\phi}\simeq
       \frac{\lambda^2 {\Phi_{0}}^2}{8\pi m_{\Phi}} < H_{0}\;. \label{GPhiphi}
       \end{equation}
       Since \mbox{$\bar{\Phi}\propto a^{-3/2}$}, it becomes
       obvious that the above constraint is the tightest in the
       early times after the amplitude of oscillations become ($
       \Phi_{0}\approx \frac{m_{\Phi}}{\sqrt\lambda}$) which takes
       place in the radiation era.
       By imposing it we get an upperbound condition for  $\lambda$ :
       \begin{equation}
       \lambda<
       \frac{m_\Phi}{M_{Pl}}
       \left(\frac{M_{Pl}}{T_{\rm eq}}\right)^{2/5} \sim 10^{-19}.
       \label{ubound}
       \end{equation}
   \item
      The condition that the oscillations of $\Phi$ are dominated
      by $V_{0}$ of eq.~\ref{V0} in the present Hubble era imply
      that ($\Phi_{0}\leq \Phi_{\Lambda}$) gets to be satisfied when
      ($\sqrt{\lambda}\left(\frac{m_{Pl}}{M_{s}} \right) > 1$)
      which , in turn, gives us a lower bound for the coupling
      constant
      \begin{equation}
       \lambda  >  10^{-30}.
      \label{lbound}
  \end{equation}
    \item
     The onset of $\Phi$-oscillations must occur in the radiation
     era $(T>1$eV) when ($ H_{osc}\sim m_{\Phi}$)
     in the aftermath of an early phase of inflation
     being followed right afterwards by reheating\cite{preh}. Their fractional
     contribution to the energy density is
     ($\frac{\rho_{\Phi}}{\rho} \propto \alpha
     \propto H^{-\frac{1}{2}}$). They eventually dominate the energy density of
     the Universe. By requiring this to take place at ($T_{eq}=1
     $eV) we find the initial displacement of $\Phi$ to be much
     smaller than the Planck scale namely
     ($\Phi_{osc} \sim 10^{-6} M_{Pl} \ll M_{Pl}$). However the
     inclusion of supergravity corrections to the potential ($
     \Delta{m_{\Phi}^{2}}\propto H(t)^{2}$)\cite{sugraRD}
      lift the flatness of the $\Phi$ direction so that $\Phi$
      begins to roll down long before ($H \sim m_{\Phi}$). Its
      motion is, however, overdamped by the excessive friction of
      a large Hubble parameter (compared to its mass) imposing a
      freeze out to the value of $\Phi$ until H is reduced enough
      for the quadratic oscillations to commence.
    \item
      Similar in spirit analysis can be applied to the study of
      the initial conditions for the Quintessence field $\Psi$
      which has to find itself near the origin ($\Psi\leq M_{s}$)in
      order to get "locked" when the $\Phi$ oscillations begin.
      The oscillations of $\Psi$ begin immediately after reheating
      with ($\Psi\propto \sqrt{\rho_{\Psi}}\propto H^{3/4}$). It
      can be analytically demonstrated that our original
      assumption for ($\Psi\approx 0$) is well justified.
      \item
      The smallness of the saddle point vacuum energy does not
      only require a small mass for our tachyonic field $\Psi$ but
      a small VEV as well ($M_{s} \sim 1 $ TeV). This can be done
      through higher order non-renormalizable terms or logarithmic
      loop corrections\cite{book}. Clearly the level of fine
      tuning implied by ($m_{\Phi} \sim 10^{15} H_{0} \sim 10^{9}
      H_{eq}$) is much less severe than the one required in most
      quintessence models $( m_{Q} \sim H_{0} )$. As a consequence and
      in contrast to quintessence models Sugra corrections in the
      matter era are negligible.
\end{itemize}
     \section{Conclusions}
      We have presented a unified model of dark matter and dark energy
      in the context of low-scale gauge-mediated supersymmetry breaking.
      Our LQCDM model retains the predictions of LCDM Cosmology,
      while avoiding
      eternal acceleration and achieving coincidence without significant
      fine-tuning. The initial conditions of our model are naturally
      attained due to the effect of supergravity corrections to the
      scalar potential in the early Universe, following a period of
      primordial inflation.
      Our oscillating $\Phi$--condensate does not have to be the dark
matter necessarily. Indeed, it is quite possible that
$\Psi$--remains locked on top of the false vacuum while
$\rho_\Phi$ is negligible at present. It is easy to see that
indeed ($ \frac{\rho_{\Phi}^{\rm min}}{\rho_{0}}\sim 10^{-30}
\lambda^{-1}$). Depending on $\lambda$, $\Phi$ may contribute only
by a small fraction to dark matter at present, while still being
able to lock quintessence and cause the observed accelerated
expansion at present. This option appears less appealing to us.

\begin{thebibliography}{99}
%
\bibitem{wmap}
D.~N.~Spergel {\it et al.},
Astrophys.\ J.\ Suppl.\  {\bf 148}, 175 (2003).
%
\bibitem{sdss}
M.~Tegmark {\it et al.}  [SDSS Collaboration],
Phys.\ Rev.\ D {\bf 69} (2004) 103501
%
\bibitem{2dF}
M.~Colless,
astro-ph/0305051.
%
\bibitem{SN}
S.~Perlmutter {\it et al.}  [Supernova Cosmology Project
Collaboration],
Astrophys.\ J.\  {\bf 517}, 565 (1999);
A.~G.~Riess {\it et al.}  [Supernova Search Team Collaboration],
Astron.\ J.\  {\bf 116}, 1009 (1998).
%
\bibitem{sahni}
G.~Jungman, M.~Kamionkowski and K.~Griest,
Phys.\ Rept.\  {\bf 267} (1996) 195
J.~R.~Primack, D.~Seckel and B.~Sadoulet,
Ann.\ Rev.\ Nucl.\ Part.\ Sci.\  {\bf 38} (1988) 751.
%
\bibitem{review}
P.~J.~E.~Peebles and B.~Ratra,
Rev.\ Mod.\ Phys.\  {\bf 75}, 559 (2003).
%
\bibitem{horizons}
S.~Hellerman, N.~Kaloper and L.~Susskind,
JHEP {\bf 0106}, 003 (2001);
W.~Fischler, A.~Kashani-Poor, R.~McNees and S.~Paban,
JHEP {\bf 0107}, 003 (2001);
E.~Witten,
hep-th/0106109;
N.~Goheer, M.~Kleban and L.~Susskind,
JHEP {\bf 0307}, 056 (2003).
%
\bibitem{quint}
L.~M.~Wang, R.~R.~Caldwell, J.~P.~Ostriker and P.~J.~Steinhardt,
Astrophys.\ J.\  {\bf 530}, 17 (2000);
I.~Zlatev, L.~M.~Wang and P.~J.~Steinhardt,
Phys.\ Rev.\ Lett.\  {\bf 82}, 896 (1999);
G.~Huey, L.~M.~Wang, R.~Dave, R.~R.~Caldwell and P.~J.~Steinhardt,
Phys.\ Rev.\ D {\bf 59}, 063005 (1999);
R.~R.~Caldwell, R.~Dave and P.~J.~Steinhardt,
Phys.\ Rev.\ Lett.\  {\bf 80}, 1582 (1998).
%
\bibitem{KL}
C.~F.~Kolda and D.~H.~Lyth,
Phys.\ Lett.\ B {\bf 458}, 197 (1999).
%
\bibitem{other}
R.~Bean and J.~Magueijo,
Phys.\ Lett.\ B {\bf 517}, 177 (2001);
L.~Amendola and D.~Tocchini-Valentini,
Phys.\ Rev.\ D {\bf 64} (2001) 043509;
D.~Tocchini-Valentini and L.~Amendola,
Phys.\ Rev.\ D {\bf 65}, 063508 (2002);
W.~Zimdahl and D.~Pavon,
Phys.\ Lett.\ B {\bf 521}, 133 (2001);
M.~Pietroni,
Phys.\ Rev.\ D {\bf 67}, 103523 (2003);
T.~Padmanabhan and T.~R.~Choudhury,
Phys.\ Rev.\ D {\bf 66}, 081301 (2002);
D.~Comelli, M.~Pietroni and A.~Riotto,
Phys.\ Lett.\ B {\bf 571}, 115 (2003);
H.~Ziaeepour,
Phys.\ Rev.\ D {\bf 69} (2004) 063512;
G.~R.~Farrar and P.~J.~E.~Peebles,
astro-ph/0307316;
E.~I.~Guendelman and A.~B.~Kaganovich,
gr-qc/0312006.
R.~G.~Cai and A.~Wang,
arXiv:hep-th/0411025.
%
\bibitem{dvali}
G.~Dvali and S.~Kachru,
hep-th/0309095;
hep-ph/0310244.
R.~Easther, J.~Khoury and K.~Schalm,
JCAP {\bf 0406} (2004) 006 [arXiv:hep-th/0402218].
%
\bibitem{ours1}
K.~Dimopoulos and M.~Axenides,
arXiv:hep-ph/0310194.
%
\bibitem{ours2}
M.~Axenides and K.~Dimopoulos,
JCAP {\bf 0407} (2004) 010
%
\bibitem{hybrid}
A.~D.~Linde,
Phys.\ Rev.\ D {\bf 49} (1994) 748
L.~Randall, M.~Soljacic and A.~H.~Guth,
Nucl.\ Phys.\ B {\bf 472} (1996) 377
G.~Lazarides,
arXiv:hep-ph/0011130.
M.~Berkooz, M.~Dine and T.~Volansky,
arXiv:hep-ph/0409226.
\bibitem{turner}
M.~S.~Turner,
Phys.\ Rev.\ D {\bf 28}, 1243 (1983)
S.~D.~H.~Hsu,
Phys.\ Lett.\ B {\bf 567} (2003) 9
%
\bibitem{book}
A.~R.~Liddle and D.~H.~Lyth, {\it Cosmological Inflation and
Large-Scale Structure} (Cambridge Univ. Press, Cambridge U.K.,
2000).
%
\bibitem{sugra}
M.~Dine, L.~Randall and S.~Thomas,
Nucl.\ Phys.\ B {\bf 458}, 291 (1996);
Phys.\ Rev.\ Lett.\  {\bf 75}, 398 (1995).
%
\bibitem{giudice}
G.~F.~Giudice and R.~Rattazzi,
Phys.\ Rept.\  {\bf 322} (1999) 419
%
\bibitem{preh}
L.~Kofman, A.~D.~Linde and A.~A.~Starobinsky,
Phys.\ Rev.\ D {\bf 56} (1997) 3258.
%
\bibitem{sugraRD}
D.~H.~Lyth and T.~Moroi,
JHEP {\bf 0405} (2004) 004.
%
\end{thebibliography}
\end{document}